# Observation of three-state nematicity in the triangular lattice antiferromagnet $Fe_{1/3}NbS_2$

Arielle Little[1,2,†], Changmin Lee[2,†], Caolan John[1,2], Spencer Doyle[1,2], Eran Maniv[1,2], Nityan L. Nair[1,2], Wenqin Chen[1,2], Dylan Rees[1,2], Jörn W.F. Venderbos[4,5], Rafael M. Fernandes[3], James G. Analytis[1,2], Joseph Orenstein[1,2,*]

[1]Department of Physics, University of California, Berkeley, California 94720, USA
[2]Materials Science Division, Lawrence Berkeley National Laboratory, Berkeley, California 94720, USA
[3]School of Physics and Astronomy, University of Minnesota, Minneapolis, Minnesota 55455, USA
[4]Department of Chemistry, University of Pennsylvania, Philadelphia, Pennsylvania 19104–6323, USA
[5]Department of Physics and Astronomy, University of Pennsylvania, Philadelphia, Pennsylvania 19104–6396, USA

[†]These authors contributed equally to this work
[*]Corresponding author (email: jworenstein@lbl.gov)

**Nematic order is the breaking of rotational symmetry in the presence of translational invariance. While originally defined in the context of liquid crystals, the concept of nematic order has arisen in crystalline matter with discrete rotational symmetry, most prominently in the tetragonal Fe-based superconductors where the parent state is four-fold symmetric. In this case the nematic director takes on only two directions, and the order parameter in such "Ising-nematic" systems is a simple scalar. Here, using a novel spatially-resolved optical polarimetry technique, we show that a qualitatively distinct nematic state arises in the triangular lattice antiferromagnet $Fe_{1/3}NbS_2$. The crucial difference is that the nematic order on the triangular lattice is a $\mathbb{Z}_3$, or three-state Potts-nematic order parameter. As a consequence, the anisotropy axes of response functions such as the resistivity tensor can be continuously re-oriented by external perturbations. This discovery provides insight into realizing devices that exploit analogies with nematic liquid crystals.**

Currently there is intense focus on incorporating antiferromagnets (AFMs) in spintronic applications, with the promise of faster response, lower threshold current, and scaling to smaller dimensions[1,2]. For example, the observation that stable switching of electric resistance can be induced by a spin-unpolarized current in thin films of antiferromagnetic CuMnAs, $Mn_2Au$, and NiO has attracted considerable attention[3-6]. In closely related work, multi-stable magnetic memory with electric read-out was demonstrated in the hexagonal AFM, MnTe[7,8]. These systems are easy-plane, collinear AFMs in which switching and memory are associated with rotation of the Néel vector, ***L***, between stable states in the plane that contains the perturbing magnetic field or current, where $\bm{L} \equiv \bm{M_1} - \bm{M_2}$, and $\bm{M_1}$, $\bm{M_2}$ are the sublattice magnetizations[9,10].

Recently, both current-induced switching and multistable memory effects were reported in the Fe-intercalated transition metal dichalcogenide (TMD) $Fe_xNbS_2$ with $x \approx 1/3$, compounds that undergo a transition to AFM order below approximately 50 K[11]. It was found that current pulses applied parallel to the TMD atomic layers could reversibly switch a 10-micron scale device between stable resistance states with significantly lower threshold current than required for CuMnAs, albeit at lower temperatures. Moreover, it was discovered that cooling through the Néel temperature ($T_N$) in an in-plane magnetic field, ***B***, induces an in-plane resistivity anisotropy whose symmetry axes continuously follow the direction of ***B*** and remain stable after the field is turned off. While the anisotropy of the static magnetic susceptibility, $\chi$, in $Fe_{1/3}NbS_2$ suggests that ***L*** is oriented primarily perpendicular to the TMD planes[12-14]. It has been suggested that this compound possesses a small in-plane component of ***L***, whose rotation by current pulses may give rise to the observed switching phenomena.

Here we report optical measurements which suggest that an in-plane nematic director plays a role in mediating switching and metastable memory in the $Fe_{1/3}NbS_2$ system. Using spatially-resolved optical polarimetry, we show that the onset of AFM order in $Fe_{1/3}NbS_2$ occurs simultaneously with a first order transition that breaks the 6-fold (screw) rotational symmetry of the paramagnetic phase. Below $T_N$ we observe three nematic domains whose directors are rotated by an angle of $2\pi/3$ with respect to each other. We provide a theoretical understanding of these results by showing that the AFM transition in $Fe_{1/3}NbS_2$ is analogous to the magneto-structural transition in Fe-based superconductors[15-19], but with a crucial difference – the $\mathbb{Z}_2$ Ising-nematic degree of freedom obtained in a tetragonal structure becomes a $\mathbb{Z}_3$ or three-state Potts-nematic on the triangular lattice[20,21].

A schematic of the optical set-up is shown in Fig. 1a. A linearly polarized probe beam is focused through a microscope objective at normal incidence onto the sample surface. The reflection amplitude from the surface is characterized by a 2 x 2 matrix, $r_{ij}$. For a material with 3-fold or higher rotational symmetry, $r_{ij} = \delta_{ij}r$, and the polarization of the probe beam remains unchanged upon reflection. Lowering rotational symmetry leads to a reflection matrix with principal optic axes *a* and *b*, and birefringence, $\Delta r \equiv r_b - r_a$. In this broken symmetry state, the probe polarization is rotated through an angle given by $\phi = Re\{r^*\Delta r \cos 2(\theta - \theta_0)\}/|r|^2$, where $\theta - \theta_0$ is the difference between the probe beam polarization and principal axis direction. By measuring $\phi(\theta)$ with a balanced optical bridge detector (see Supplementary Information) we obtain the optic axes directions *a* and *b* and birefringence amplitude, $\Delta r$.

To obtain the maps of $\Delta r$ and $\theta_0$ described below we overlap the probe beam with an 800 nm pump laser chopped at 2 kHz, which modulates the sample temperature. Lock-in detection at the pump chopping frequency eliminates long-time-period drifts and enables microradian sensitivity to the modulated polarization rotation, $\delta\phi$. Fig. 1b shows $\delta\phi$ as a function of temperature, $T$, at a single 10-micron spot on the sample and for a fixed probe polarization. The abrupt onset of birefringence at 49 K indicates a first-order phase transition to a state in which the 6-fold screw symmetry of the paramagnetic state is lowered to at most 2-fold rotational symmetry. The decaying oscillations in $\delta\phi$ with further lowering of $T$ are explained below. Plotted in Fig. 1c is $\delta\phi$ as a function of probe polarization angle $\theta$ in the vicinity of the phase transition. Below $T_N$, $\delta\phi$ shows the expected $\cos 2(\theta - \theta_0)$ dependence, allowing us to determine the orientation of the optic axes in the magnetic phase.

The clue as to the origin of the oscillations in $\delta\phi(T)$ shown in Fig. 1b is that the optical probe detects the onset of order at a higher temperature than bulk probes do. Figure 2a shows a summary of bulk measurements in the temperature range of the transition. A singularity in heat capacity, $C_p$, coincides with abrupt jumps in $\chi_\perp$ and in-plane resistivity, $\rho_{xx}$, indicating a single, first-order transition to the AFM state at 43 K.

The oscillatory $\delta\phi(T)$ is plotted in conjunction with $C_p$ in Fig. 2b, showing that there is a large offset between the transition temperatures measured optically (49 K) and by bulk probes (43 K).

Figure 2c shows the polarization rotation $\phi(T)$ measured directly, that is, without photo-thermal modulation. The presence of the same oscillations in $\phi(T)$ seen in $\delta\phi(T)$ confirms that they are not an artifact of photo-thermal modulation. Instead, our analysis, presented in detail in the Supplementary Information, shows that oscillations appear in the optical measurements because $T_N$ decreases with increasing depth below the surface. For temperatures below the surface transition temperature $T_N(0) = 49$ K, a buried interface separates a birefringent surface layer from the isotropic bulk, as illustrated in Fig. 2d. As the depth, $z$, of the interface increases with decreasing $T$, the phase difference between the reflections from the surface and the buried interface produces the interference pattern seen in $\phi(T)$. The oscillations are cut off when the buried interface reaches beyond the penetration depth of the probe beam, which is approximately 150 nm[22]. At this depth, $T_N(z)$ has reached 47.5 K, which is still 4.5 K higher than the bulk $T_N$. The mesoscopic scale of the probe wavelength in the medium enables a new method for depth profiling of transition temperatures on a 10-nm scale.

We now turn to the spatial mapping of the amplitude and principal axis directions of the nematic order. Below the transition we detect three orientations of optic axes offset by $2\pi/3$ from each other, as illustrated by polar plots of $\delta\phi(\theta)$ at three locations on the sample (Fig. 3a). By registration of the x-ray Laue diffraction pattern with the probe polarization angle, we find that the three orientations of optic axes correspond to the three crystallographic symmetry directions of the triangular Fe-lattice (Fig. 4b). We assign a color to each of the orientations, such that [100] is red, [010] is green, and [$\overline{1}10$] is blue. A map of a 900 $\mu$m × 500 $\mu$m region of the sample is shown in Fig. 3b, revealing the presence of all three domains, which can be as large as hundreds of microns or as small as our resolution of 10 microns. The domain distribution is deterministic upon warming and cooling, and does not change in magnetic fields up to 400 G.

The nematic order that accompanies the onset of AFM order in Fe$_{1/3}$NbS$_2$ can be qualitatively understood to result from the geometric frustration of Ising spins[23-25] on the triangular ($\sqrt{3}a \times \sqrt{3}a$) superlattice[26-29] of Fe atoms. A distortion along one direction of Fe-Fe bonds relieves the frustration, allowing one of three degenerate ordered phases to condense. Both stripe-like order (ordering wavevectors $\boldsymbol{Q}_i$ in the $\Gamma - M$ directions) and zigzag order ($\boldsymbol{Q}_i$ in the $\Gamma - K$ directions) are possible (Figs. 4b,c), and single-crystal neutron measurements are necessary to fully distinguish between the two[12]. Both types of order give rise to nematicity and the stripe case is discussed here (see Supplementary Information for zigzag).

To gain further insight into the nature of the transition, we construct a phenomenological Landau model assuming stripe order. We introduce three order parameters $\boldsymbol{L}_1$, $\boldsymbol{L}_2$, and $\boldsymbol{L}_3$, whose wavevectors are parallel to one of the three $\Gamma - M$ directions. Spin-rotational and 6-fold rotational symmetries restrict the corresponding magnetic Landau free-energy $F_M$ to the form,

$$F_M = a \sum_{i=1,3} L_i^2 + v_0 \left( \sum_{i=1,3} L_i^2 \right)^2 + v_1 \sum_{i<j} L_i^2 L_j^2 + v_2 \sum_{i<j} (\boldsymbol{L}_i \cdot \boldsymbol{L}_j)^2. \quad (1)$$

For simplicity, here we assume commensurate wave-vectors (Fig. 4c), but the results can be extended in a straightforward way to incommensurate order. The Landau coefficients $v_1$ and $v_2$ determine the nature of the magnetic state below the Néel temperature. In particular, for $v_1 > 0$ and $v_2 > -v_1$ a single-$Q$ magnetic state is favored in which only one of the $\boldsymbol{L}_i$ is nonzero and rotational symmetry is broken. In contrast, other parameter values lead to equal amplitude triple-$Q$ magnetic ordering, preserving the rotational symmetry of the paramagnetic state. While the particular values of these and the other Landau coefficients $a$ and $v_0$ depend on microscopic considerations related to the mechanism responsible for the magnetic instability, for our purposes a phenomenological approach suffices.

Indeed, our experimental observation of rotational symmetry breaking points uniquely to single-$Q$ AFM order in Fe$_{1/3}$NbS$_2$ (i.e. $v_1 > 0$ and $v_2 > -v_1$), in which $\langle \boldsymbol{L}_i \rangle \neq 0$ for one of the three order parameter components. As seen in Fig. 4d, the six nearest-neighbor links are no longer equivalent in the stripe AFM state; four bonds couple anti-parallel spins whereas two bonds couple parallel spins. This rotational symmetry breaking is captured by an order parameter $\boldsymbol{n} = (n_1, n_2)$, which is given in terms of the magnetic components as,

$$(n_1, n_2) = \left(\boldsymbol{L}_1^2 + \boldsymbol{L}_2^2 - 2\boldsymbol{L}_3^2, \sqrt{3}\boldsymbol{L}_1^2 - \sqrt{3}\boldsymbol{L}_2^2\right). \tag{2}$$

Here $\boldsymbol{n}$ is a nematic director indicative of rotational symmetry breaking and can be parametrized as $\boldsymbol{n} = n(\cos 2\theta, \sin 2\theta)$. The structure of $\boldsymbol{n}$ for the triangular lattice contrasts sharply with nematic order in tetragonal systems such as the Fe-pnictides, which is described by a single-component $\mathbb{Z}_2$ Ising-nematic order parameter[15,18].

The magnetic free energy (Eq. 1) can be used to derive an effective free energy $F_n$ for nematic order, since the nematic order parameter (Eq. 2) is a composite magnetic order parameter. Integrating out magnetic fluctuations and going beyond mean-field theory, one finds,

$$F_n = \alpha n^2 + \beta n^3 \cos 6\theta + \gamma n^4, \tag{3}$$

where the Landau parameters $\alpha$, $\beta$, and $\gamma$ depend on the Landau parameters of the magnetic free energy, Eq. 1. The key observation is that Eq. 3 represents the same free energy as that of the three-state Potts model[21,30], implying that $\boldsymbol{n}$ is a $\mathbb{Z}_3$ Potts-nematic order parameter. This follows from the third-order term in Eq. 3, which defines a $\mathbb{Z}_3$ Potts-anisotropy and restricts the director to point along one of the three high-symmetry directions of the lattice. The first-order jump in the nematic order parameter arising from the cubic term triggers a jump in the magnetic correlation length and a simultaneous first-order nematic-AFM transition consistent with the data.

Next we show that the relative population of the three AFM domains can be tuned in response to an external perturbation that couples to the nematic director, in this case uniaxial strain. The birefringence map shown in Fig. 3b was obtained with the sample resting on a Cu plate secured by vacuum grease, a configuration in which the external strain from thermal contraction is negligibly small. The inset presents a histogram illustrating the domain population measured in this configuration, showing that all three domains are represented.

To apply strain the sample was glued onto a piezoelectric stack, which upon cooling applies a uniaxial strain of ≈ 0.1 %[31-33]. Imaging the same region of the sample with compressive strain applied parallel to the bond direction, as illustrated in Fig. 5a, yields the birefringence map shown in Fig. 5b. This map, together with the domain histogram (Fig. 5c), shows that the areal fraction of the red [100] domains, whose principal axes are parallel and perpendicular to the bond direction, is strongly suppressed and redistributed to green and blue domains. The sample was then cleaved, rotated by 90° and re-mounted on the piezo stack such that now tensile strain is applied in the bond direction (Fig. 5d). In this configuration the areal fraction of the red domains grows at the expense of the other two (Figs. 5e,f). In the Supplementary Information we show that the domain repopulation follows in a straightforward fashion from the coupling of the strain tensor to the nematic director, ***n***. Basically, tensile strain acts as a positive nematic conjugate field oriented parallel to the strain direction, favoring the domain corresponding to this direction. On the other hand, compressive strain acts as a negative conjugate field, which suppresses the corresponding domain and favors the other two domains.

The repopulation of $\mathbb{Z}_3$ nematic domains in response to uniaxial strain suggests a mechanism for switching and memory phenomena in response to in-plane perturbations ***B*** and currents, ***J***. In $Fe_{1/3}NbS_2$, ***n*** can couple to such perturbing fields, playing the same role as the in-plane ***L*** does in the easy-plane systems such as CuMnAs and MnTe. As shown for coupling to ***L*** in MnTe[6], and for coupling to ***n*** in $Fe_{1/3}NbS_2$ (see Supplementary Information), application of in-plane ***B*** or ***J*** unbalances the domain population, resulting in a global resistivity tensor whose principal axes continuously follow[34] the direction of ***B*** or ***J***. We note that this effect cannot occur in a $\mathbb{Z}_2$ Ising-nematic system, where the anisotropy is locked to the crystal axes, regardless of the direction of the perturbation[15,32].

In conclusion, we have shown that the onset of AFM order in $Fe_{1/3}NbS_2$ occurs via a first-order transition that lowers the rotational symmetry of the triangular lattice from 6 to at most 2-fold. Below $T_N$, maps of local birefringence reveal three domains, characterized by in-plane nematic directors with relative angles $2\pi/3$. The first-order nature of this transition may be understood in terms of a three-state Potts model where the symmetry-breaking order parameter is a $\mathbb{Z}_3$ nematic, which in turn is a composite magnetic order parameter. Furthermore, we demonstrated that the relative population of $\mathbb{Z}_3$ nematic domains can be controlled by uniaxial strain. In principle, the coupling of external perturbations to a nematic director could be sufficiently strong to enable low switching thresholds in nematic-magnetic systems.

**Methods**

Imaging the direction of the optic axes is performed by mounting the sample on a cryogenic piezo-actuated *xyz* stage in a Montana Instruments 4 K cryostat with a low-working-distance cryogenic window. As the sample is rastered beneath the probe beam focus, a polarization scan of *δϕ* is taken at each position on a grid. Polarization scanning is enabled by co-rotating half wave plates, a scheme described in more detail in the Supplementary Information. Whereas dichroism-based imaging[35,36] is limited to the contrast between two fixed orthogonal polarizations, here the continuous control of the probe polarization measures the precise local orientation of the optic axes.

**Acknowledgements:**

We thank D.H. Lee and J. Moore for useful discussions and N. Tamura for support at the Advanced Light Source. Optical measurements were performed at the Lawrence Berkeley National Laboratory in the Quantum Materials program supported by the supported by the Director, Office of Science, Office of Basic Energy Sciences, Materials Sciences and Engineering Division, of the U.S. Department of Energy under Contract No. DE-AC02- 05CH11231. A.L. and J.O. received support for optical measurements from the Gordon and Betty Moore Foundation's EPiQS Initiative through Grant No. GBMF4537 to J. O. at UC Berkeley. Synthesis of $Fe_{1/3}NbS_2$ was supported by Laboratory Directed Research and Development Program of Lawrence Berkeley National Laboratory under Contract No. DE-AC02-05CH11231. J.G.A. and N.L.N. received support from the Gordon and Betty Moore Foundation's EPiQS Initiative Grant No. GBMF4374 to J. G.A. at UC Berkeley. R.M.F. is supported by the U.S. Department of Energy, Office of Science, Basic Energy Sciences, under Award DE-SC0012336. X-ray diffraction to register crystal orientation was carried out at beam line 12.3.2 at the

**Author contributions:**

A.L. and C.L. performed and contributed equally to the birefringence microscopy measurements and data analysis. C.J., S.D. and E.M. grew and characterized the crystals. E.M., N.L.N., and J.G.A. discovered the switching effect which motivated this project. J.W.F.V. and R.M.F. developed the theoretical model. W.C. performed the simulation of depth profiling. D.R. assisted with optical measurements. J.O., A.L., C.L., J.W.F.V., and R.M.F. wrote the manuscript. All authors commented on the manuscript.

**Competing Interests:**

The authors declare no competing financial interests.

**References**

1	Jungwirth, T., Marti, X., Wadley, P. & Wunderlich, J. Antiferromagnetic spintronics. *Nat. Nanotechnol.* **11**, 231-241, (2016).

2	Baltz, V. *et al.* Antiferromagnetic spintronics. *Rev. Mod. Phys.* **90**, 015005, (2018).

3	Wadley, P. *et al.* Electrical switching of an antiferromagnet. *Science* **351**, 587-590, (2016).

4	Bodnar, S. Y. *et al.* Writing and reading antiferromagnetic $Mn_2Au$ by Néel spin-orbit torques and large anisotropic magnetoresistance. *Nat. Commun.* **9**, 348, (2018).

5	Moriyama, T., Oda, K., Ohkochi, T., Kimata, M. & Ono, T. Spin torque control of antiferromagnetic moments in NiO. *Sci. Rep.* **8**, 14167, (2018).

6	Saidl, V. *et al.* Optical determination of the Néel vector in a CuMnAs thin-film antiferromagnet. *Nat. Photon.* **11**, 91, (2017).

7	Kriegner, D. *et al.* Multiple-stable anisotropic magnetoresistance memory in antiferromagnetic MnTe. *Nat. Commun.* **7**, 11623, (2016).

8	Kriegner, D. *et al.* Magnetic anisotropy in antiferromagnetic hexagonal MnTe. *Phys. Rev. B* **96**, 214418, (2017).

9   Železný, J. *et al.* Relativistic Néel-Order Fields Induced by Electrical Current in Antiferromagnets. *Phys. Rev. Lett.* **113**, 157201, (2014).

10  Železný, J. *et al.* Spin-orbit torques in locally and globally noncentrosymmetric crystals: Antiferromagnets and ferromagnets. *Phys. Rev. B* **95**, 014403, (2017).

11  Nair, N. L. *et al.* Electrical switching in a magnetically intercalated transition metal dichalcogenide, Preprint at: https://arxiv.org/abs/1907.11698> (2019).

12  Van Laar, B., Rietveld, H. M. & Ijdo, D. J. W. Magnetic and crystallographic structures of $Me_xNbS_2$ and $Me_xTaS_2$. *J. Solid State Chem.* **3**, 154-160, (1971).

13  Gorochov, O., Blanc-soreau, A. L., Rouxel, J., Imbert, P. & Jehanno, G. Transport properties, magnetic susceptibility and Mössbauer spectroscopy of $Fe_{0.25}NbS_2$ and $Fe_{0.33}NbS_2$. *Philos. Mag. B* **43**, 621-634, (1981).

14  Yamamura, Y. *et al.* Heat capacity and phase transition of $Fe_xNbS_2$ at low temperature. *J. Alloys Compd.* **383**, 338-341, (2004).

15  Chu, J.-H. *et al.* In-Plane Resistivity Anisotropy in an Underdoped Iron Arsenide Superconductor. *Science* **329**, 824-826, (2010).

16  Johnston, D. C. The puzzle of high temperature superconductivity in layered iron pnictides and chalcogenides. *Adv. Phys.* **59**, 803-1061, (2010).

17  Paglione, J. & Greene, R. L. High-temperature superconductivity in iron-based materials. *Nat. Phys.* **6**, 645, (2010).

18  Fernandes, R. M., Chubukov, A. V. & Schmalian, J. What drives nematic order in iron-based superconductors? *Nat. Phys.* **10**, 97, (2014).

19  Si, Q., Yu, R. & Abrahams, E. High-temperature superconductivity in iron pnictides and chalcogenides. *Nat. Rev. Mater.* **1**, 16017, (2016).

20  Fernandes, R. M., Orth, P. P. & Schmalian, J. Intertwined Vestigial Order in Quantum Materials: Nematicity and Beyond. *Annu. Rev. Condens. Matter Phys.* **10**, 133-154, (2019).

21  Hecker, M. & Schmalian, J. Vestigial nematic order and superconductivity in the doped topological insulator $Cu_xBi_2Se_3$. *npj Quantum Mater.* **3**, 26, (2018).

22  Fan, S. *et al.* Electronic chirality in the metallic ferromagnet $Fe_{1/3}TaS_2$. *Phys. Rev. B* **96**, 205119, (2017).

23  Wannier, G. H. Antiferromagnetism. The Triangular Ising Net. *Phys. Rev.* **79**, 357-364, (1950).

24  Korshunov, S. E. Nature of phase transitions in the striped phase of a triangular-lattice Ising antiferromagnet. *Phys. Rev. B* **72**, 144417, (2005).

25  Smerald, A., Korshunov, S. & Mila, F. Topological Aspects of Symmetry Breaking in Triangular-Lattice Ising Antiferromagnets. *Phys. Rev. Lett.* **116**, 197201, (2016).

26  Friend, R. H., Beal, A. R. & Yoffe, A. D. Electrical and magnetic properties of some first row transition metal intercalates of niobium disulphide. *Philos. Mag. A* **35**, 1269-1287, (1977).

27  Parkin, S. S. P. & Friend, R. H. 3d transition-metal intercalates of the niobium and tantalum dichalcogenides. I. Magnetic properties. *Philos. Mag. B* **41**, 65-93, (1980).

28  Horibe, Y. *et al.* Color theorems, chiral domain topology, and magnetic properties of Fe(x)TaS2. *J Am Chem Soc* **136**, 8368-8373, (2014).

29  Doyle, S. *et al.* Tunable Giant Exchange Bias in an Intercalated Transition Metal Dichalcogenide, Preprint at: https://arxiv.org/abs/1904.05872 (2019).

30  Straley, J. P. & Fisher, M. E. Three-state Potts model and anomalous tricritical points. *J. Phys. A* **6**, 1310-1326, (1973).

31  Simpson, A. M. & Wolfs, W. Thermal expansion and piezoelectric response of PZT Channel 5800 for use in low-temperature scanning tunneling microscope designs. *Rev. Sci. Instrum.* **58**, 2193-2195, (1987).

32  Chu, J.-H., Kuo, H.-H., Analytis, J. G. & Fisher, I. R. Divergent Nematic Susceptibility in an Iron Arsenide Superconductor. *Science* **337**, 710-712, (2012).

33  Hicks, C. W., Barber, M. E., Edkins, S. D., Brodsky, D. O. & Mackenzie, A. P. Piezoelectric-based apparatus for strain tuning. *Rev. Sci. Instrum.* **85**, 065003, (2014).

34  Fradkin, E., Kivelson, S. A., Lawler, M. J., Eisenstein, J. P. & Mackenzie, A. P. Nematic Fermi Fluids in Condensed Matter Physics. *Annu. Rev. Condens. Matter Phys.* **1**, 153-178, (2010).

35  Sapozhnik, A. A. *et al.* Direct imaging of antiferromagnetic domains in $Mn_2Au$ manipulated by high magnetic fields. *Phys. Rev. B* **97**, 134429, (2018).

36  Bodnar, S. Y. *et al.* Imaging of current induced Néel vector switching in antiferromagnetic $Mn_2Au$. *Phys. Rev. B* **99**, 140409, (2019).

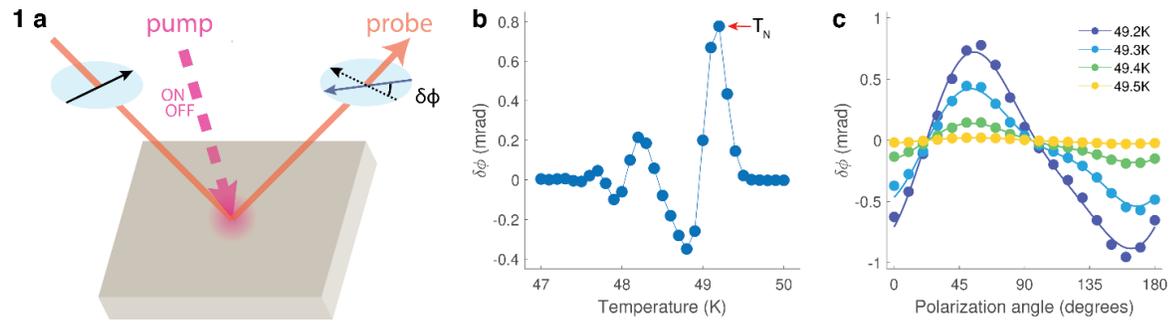

**Fig. 1 | Photo-thermal modulated birefringence measurements. a,** Pump and probe beams are spatially overlapped onto the sample surface. The 633 nm beam probes the thermally modulated polarization rotation induced by 24 $\mu$W of 800 nm pump that is modulated at 2 kHz with an optical chopper. The dependence on probe beam polarization is measured using a pair of co-rotating half-wave plates. **b**, Photo-thermally induced polarization rotation $\delta\phi$ exhibits a sharp peak at $T_N$ followed by decaying oscillations. **c**, Polarization rotation $\delta\phi$ as a function of input probe polarization angle at temperatures close to $T_N$. The input polarization angle with maximal $\delta\phi$ indicates the direction of the optic fast axis.

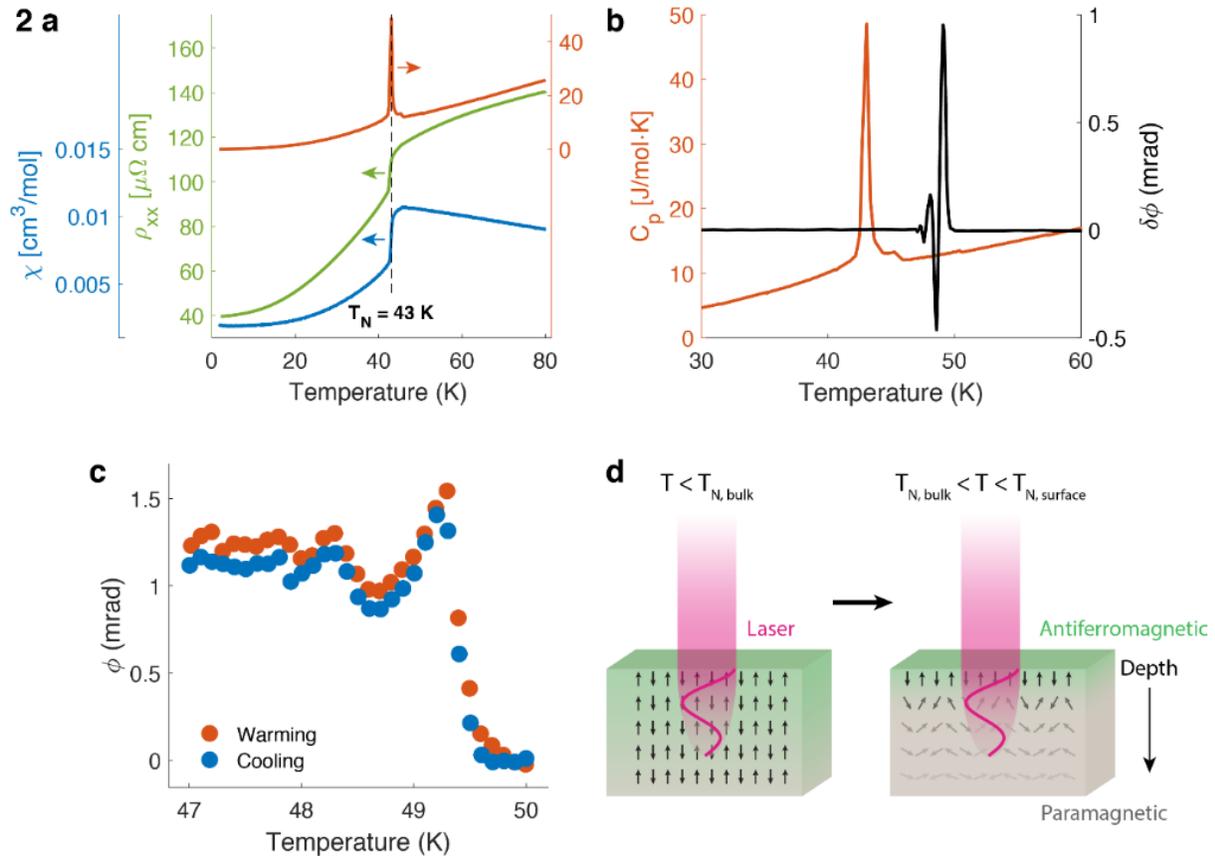

**Fig. 2 | First-order phase transitions at surface and bulk. a,** Heat capacity (red), resistivity (green), and magnetic susceptibility (blue) measurements all reveal a sharp first-order phase transition at $T_N$ = 43 K. **b,** Optical birefringence data (black) shows a higher $T_N$ (= 49 K) compared to heat capacity (red) measurement. **c,** Unmodulated birefringence data also exhibits oscillations at temperatures slightly below the surface $T_N$. **d,** Illustration of surface and bulk in two temperature regimes, showing the presence of a buried interface when $T$ is between the surface and bulk transition temperatures.

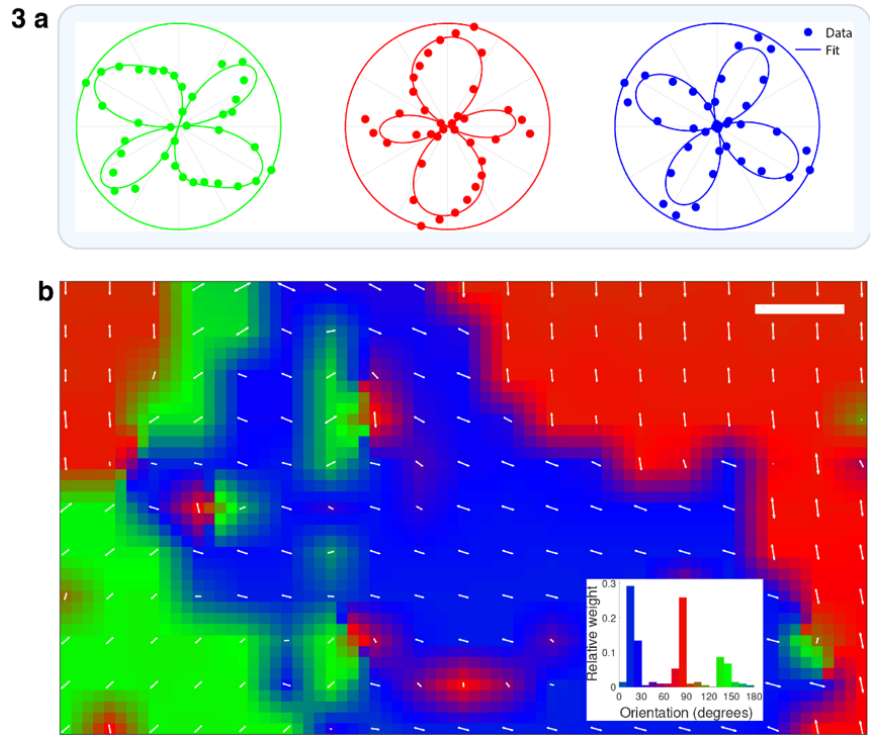

**Fig. 3 | Crystal structure and birefringence map. a,** Polar plots of $\delta\phi$ at different sample locations reveal breaking of rotational symmetry. **b**, Birefringence map across a 900 $\mu$m × 500 $\mu$m area of the sample shows three distinct domains whose optic fast axes are offset by 120° from one another. This region was sampled in 50 um steps, and the positions of the foci are indicated by the white arrows. The three nematic domains in **c** are color-coded as red, green, and blue. Each is associated with a high-symmetry direction of the Fe-Fe triangular superlattice, as shown in Fig. 4b. Scale bar: 100 μm. Inset: histogram showing the distribution of domain population in the birefringence map.

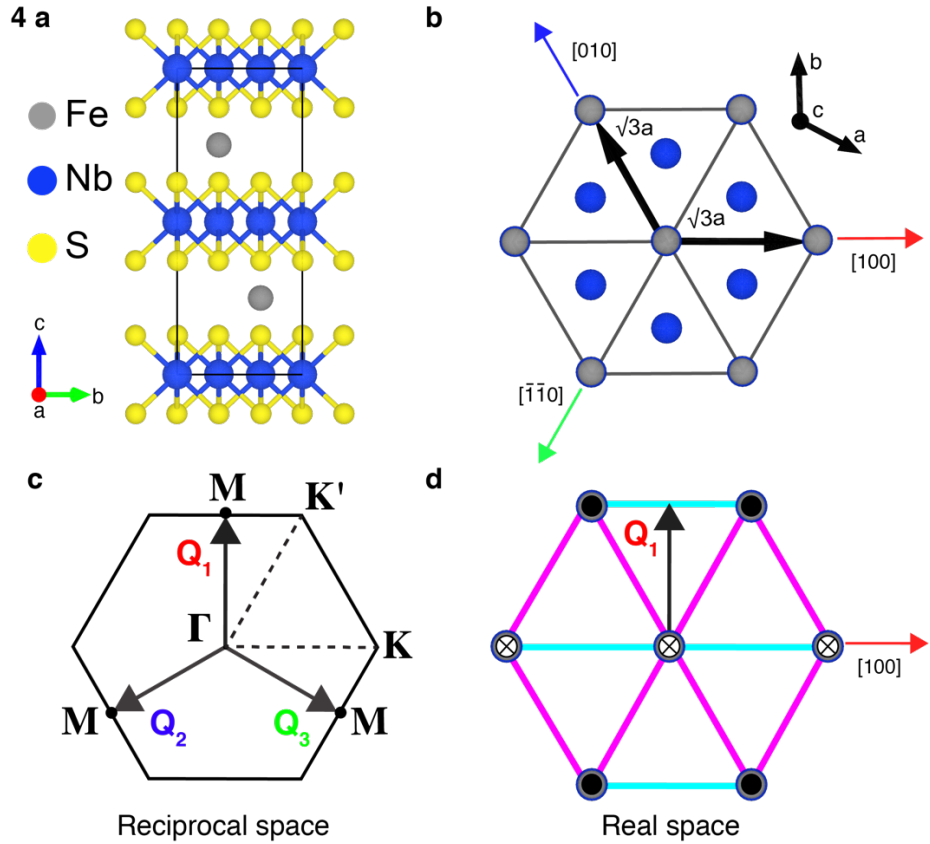

**Fig 4| Crystal structure and nematic order. a,** Cut along the c-direction of $Fe_{1/3}NbS_2$ illustrating staggered intercalation of Fe-ions between the TMD planes. **b,** View of triangular Fe-superlattice in the *ab* plane. Colored arrows indicate the high-symmetry directions and three domain orientations shown in Fig. 3. **c,** Hexagonal Brillouin zone for a single Fe-layer with high-symmetry directions shown. The three $\Gamma - M$ ordering wave-vectors ($\mathbf{Q}_1$, $\mathbf{Q}_2$, $\mathbf{Q}_3$), related by $2\pi/3$ rotations, corresponding to the three domains. **d,** Real space cartoon of $\mathbf{Q}_1$ stripe order. Solid black circles indicate spin up, white circles with black X's indicate spin down. Light blue bonds connect parallel spins, pink bonds connect anti-parallel spins.

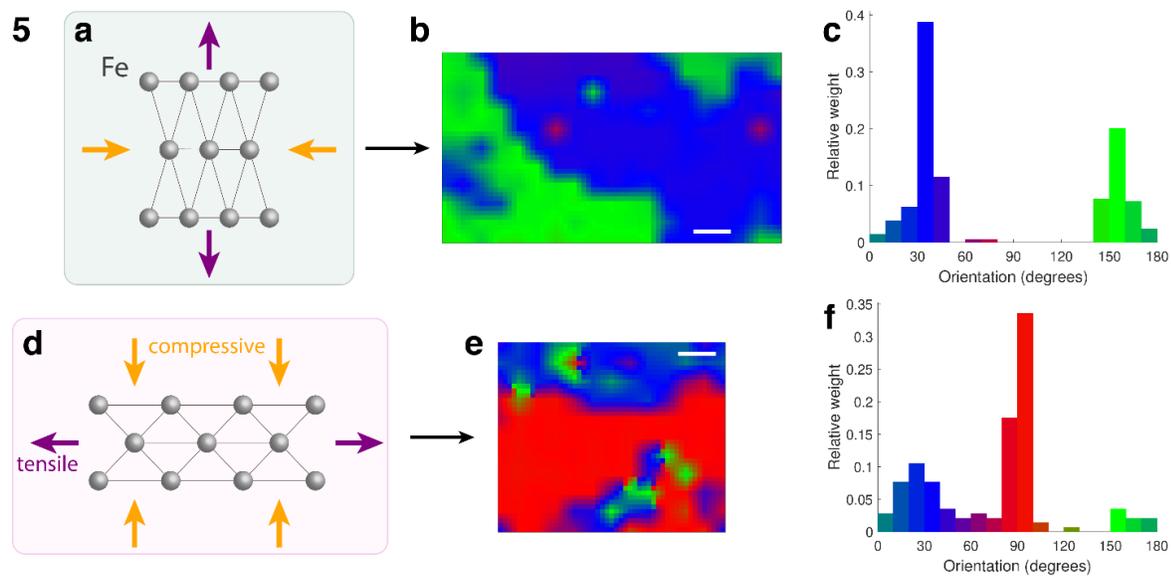

**Fig. 5 | Strain tuning of birefringent domains. a,** Compressive strain is applied along the [100] direction of the lattice, parallel to the Fe-Fe bond direction. **b,** The birefringence map and **c,** histogram showing that the green and blue domains are favored over the red. **d,** Conversely, tensile strain along the [100] direction results in dominance of red domains, as illustrated by the **(e)** birefringence map and **(f)** histogram.

# Supplementary Information
"Observation of three-state nematicity in the triangular lattice antiferromagnet Fe$_{1/3}$NbS$_2$"

## I. Sample Growth

Single crystals of Fe$_{1/3}$NbS$_2$ were grown using vapor transport techniques.[1] The samples studied are hexagonal crystals 1-3 mm in size and 20 – 100 microns thick. Samples were grown with a nominal Fe-concentration of $x$ = 0.37. Energy-dispersive x-ray spectroscopy (EDX) and inductively coupled plasma (ICP) measurements were performed to verify the chemical composition, yielding an Fe-intercalation value of $x$ = 0.34. X-ray diffraction and transmission electron microscopy measurements[1] confirm the crystal structure is the nominal $x$ =1/3 non-centrosymmetric space group $P6_3 22$.

## II. Birefringence Microscopy and Balanced Optical Bridge Detector

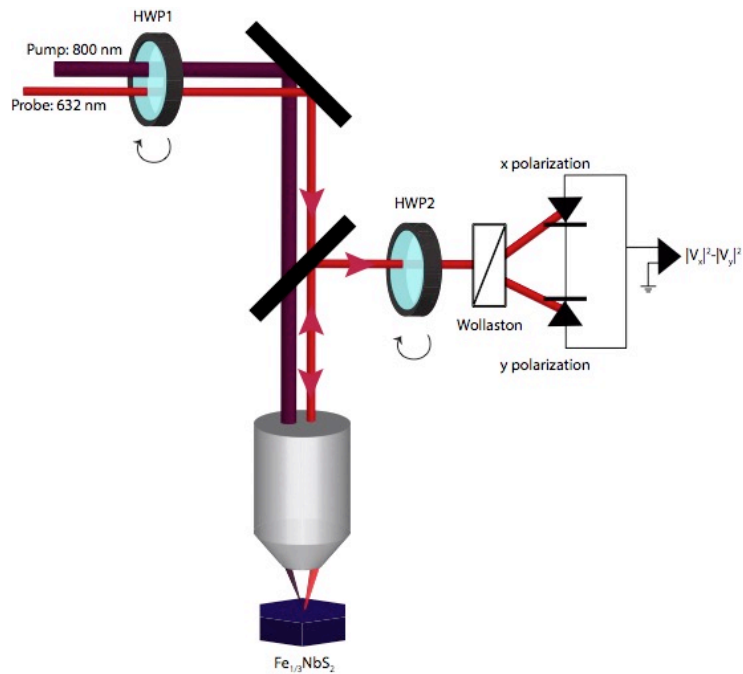

**Fig. S1** | Schematic of photo-thermal modulated birefringence microscopy and optical bridge detector.

A schematic of the measurement apparatus is shown in Fig. S1. The polarization of the incident probe laser is set by a rotatable half-wave plate, HWP1. For the photothermal modulation measurements, the pump is modulated at 2 KHz by an optical chopper and both beams are focused onto the sample surface with a microscope objective. The measured $\delta\phi(T)$ is independent of the pump polarization. The pump reflection is rejected via a color filter while the reflected probe is sent through a Wollaston prism which spatially separates the horizontal ($x$) and vertical ($y$) polarization components of the beam. The two orthogonally polarized beams are sent to an optical bridge detector consisting of two unbiased photodiodes connected in parallel but with opposite polarity. The net photocurrent will be zero, or 'balanced', when the $x$ and $y$ components have equal intensity. Balancing is achieved by warming the sample above the Néel temperature and adjusting a second half-wave plate (HWP2).

In the antiferromagnetic (AFM) phase the sample becomes birefringent, rotating the polarization of the reflected probe and unbalancing the detector. The intensity admitted to each of the two photodiodes can be calculated using Jones calculus,

$$\begin{bmatrix} V_x \\ V_y \end{bmatrix} = \begin{bmatrix} \cos\theta & -\sin\theta \\ \sin\theta & \cos\theta \end{bmatrix} \begin{bmatrix} r_a & 0 \\ 0 & r_b \end{bmatrix} \begin{bmatrix} \cos\theta & \sin\theta \\ -\sin\theta & \cos\theta \end{bmatrix} \begin{bmatrix} 1 \\ 1 \end{bmatrix}, \quad (1)$$

where $\theta$, set by the two half-wave plates, represents the angle between the incident probe beam polarization and the sample principal optic axes, $\boldsymbol{a}$ and $\boldsymbol{b}$. The reflection coefficients along each of the sample principle axes are written as $r_a$ and $r_b$. The balanced diode detector measures the intensity difference between the two polarization components, $|V_x|^2 - |V_y|^2$. To first order in $\Delta r/\bar{r}$, where,

$$\bar{r} = \frac{r_a + r_b}{2}, \quad \Delta r = r_a - r_b, \quad (2)$$

we obtain,

$$|V_1|^2 - |V_2|^2 = 2Re(\bar{r}^*\Delta r)\cos 2\theta. \quad (3)$$

To measure the unmodulated polarization rotation, $\phi(T)$, for a fixed HWP1 angle as in the main text Fig. 2c, the pump is removed entirely. The optical chopper is moved to modulated the probe beam and detection method remains the same.

## III. Depth Profiling of the AFM Transition

As discussed in the main text, the AFM phase sets in at the sample surface at a higher temperature than in the bulk. The temperature oscillations in $\phi(T)$ shown in the main text Fig 2(b), (c) are well-described by a model wherein a birefringent surface layer propagates into the bulk with decreasing temperature.

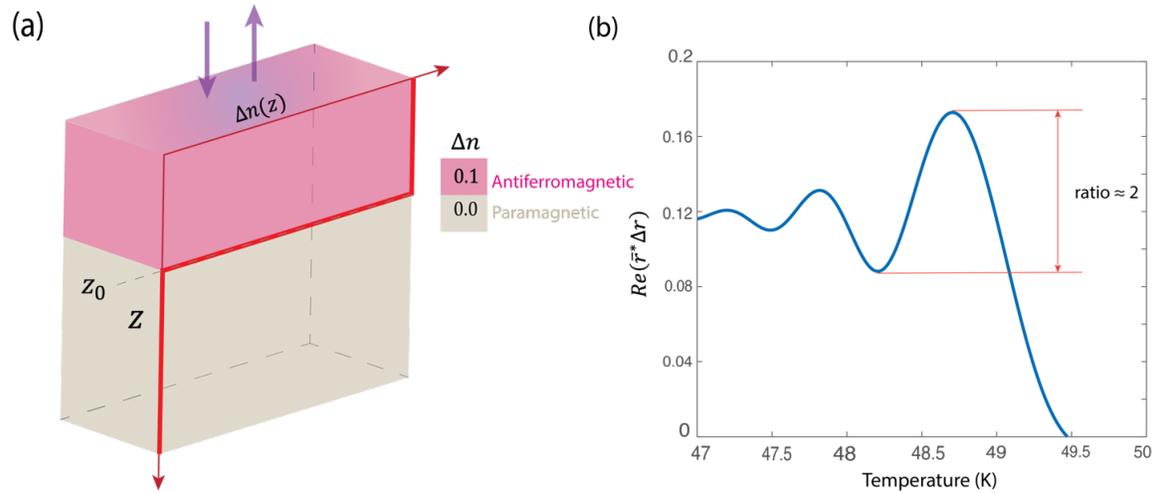

**Fig. S2 | a,** Illustration of the sharp, shifting interface between the AFM phase, where $\Delta n = 0.1$, and the paramagnetic phase, where $\Delta n = 0$. **b,** The result of simulation. The ratio between the first maximum point and the first minimum point is approximately 2.

We first consider a sharp interface between the AFM region and the paramagnetic region, as illustrated in Fig. S2a. In the paramagnetic region we assume an isotropic index of refraction given by $= 2.6 + 0.6i$ , the value measured for the similar compound $Fe_{1/4}TaS_2$ by Fan et al[2]. The AFM surface region has a birefringence characterized by the difference in index of refraction along the optic axes, $\Delta n = n_a - n_b$. The simulation result shown in Fig. S2b assumes a relative index birefringence of 0.04 and confirms the sharp interface model provides a good qualitative explanation of the data. However, for this simple case the oscillation amplitude is larger than in the data shown in main text Fig. 2c. In the measurement, the ratio between the first maximum point and the first minimum point is approximately 1.5, but this sharp interface calculation gives a ratio of 2.

This discrepancy may be remedied by considering a smooth interface with a finite width. We model the smooth interface by discretizing the interface into multiple layers, or 'slabs', with varying $\Delta n$, as illustrated in Fig. S3a. We use a modified Fermi-Dirac distribution function to describe the broadened interface,

$$\Delta n(z) = \frac{\Delta n_1}{e^{\frac{z-z_0}{\alpha}} + 1}. \tag{4}$$

with three parameters: $\Delta n_1$, the difference between refractive indices along two principal optical axes in the ordered phase; $z_0$, the depth of interface as a function of temperature; and $\alpha$, the interface width.

Our simulation is carried out using the scattering matrix method[3], which allows us to calculate the transmitted and reflected electric field through a series of birefringent slabs. First, we calculate the $I_i$ matrices that describe transmitted electric field amplitude at the surface of a given slab and the $T_i$ matrices that describe the phase shift of the electric field as it is transmitted through the slab,

$$I_i = \begin{bmatrix} \frac{1}{t_i} & \frac{r_i}{t_i} \\ \frac{r_i}{t_i} & \frac{1}{t_i} \end{bmatrix}, \qquad T_i = \begin{bmatrix} e^{i\delta_i} & 0 \\ 0 & e^{-i\delta_i} \end{bmatrix}, \tag{5}$$

where $r_i$ is the reflection coefficient and $t_i$ is the transmission coefficient for the $i^{\text{th}}$ slab. The phase shift over a propagation length $\Delta z_i$ is given by $\delta_i = 2\pi n_i \Delta z_i / \lambda$. Second, we calculate the scattering matrix by multiplying all the $I_i$ and $T_i$ matrices in sequence,

$$M = I_o T_1 I_1 T_2 I_2 \ldots I_{m-1} T_m I_m. \tag{6}$$

We can now calculate $E^{\text{in}}$ and $E^{\text{out}}$ at the sample surface separately for the polarization components along the principle optic axes,

$$\begin{bmatrix} E_a^{\text{in}} \\ E_a^{\text{out}} \end{bmatrix} = M_a \begin{bmatrix} 1 \\ 0 \end{bmatrix}, \qquad \begin{bmatrix} E_b^{\text{in}} \\ E_b^{\text{out}} \end{bmatrix} = M_b \begin{bmatrix} 1 \\ 0 \end{bmatrix}. \tag{7}$$

Finally, we can calculate the detector output, $Re(\bar{r}^* \Delta r)$,

$$r_a = \frac{E_a^{\text{out}}}{E_a^{\text{in}}}, \qquad r_b = \frac{E_b^{\text{out}}}{E_b^{\text{in}}}, \tag{8}$$

$$Re(\bar{r}^* \Delta r) = Re\left[\left(\frac{r_a + r_b}{2}\right)^* (r_a - r_b)\right]. \tag{9}$$

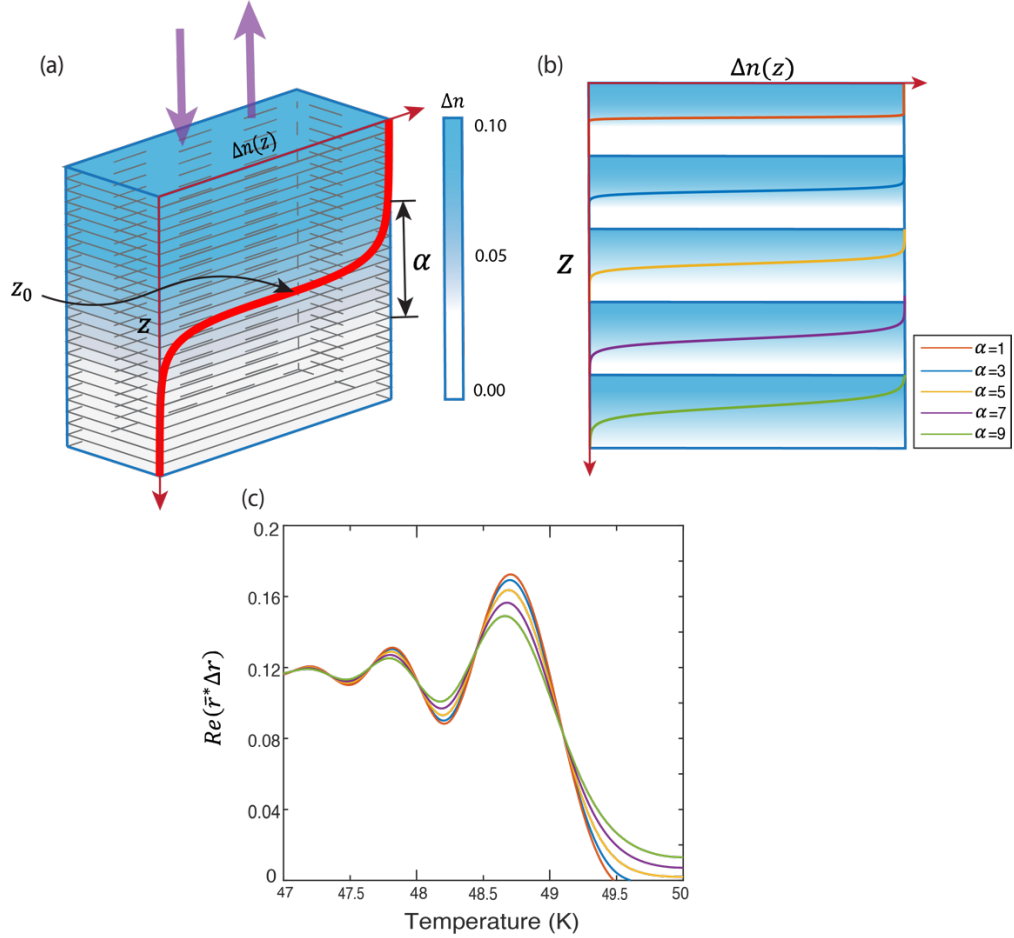

**Fig. S3 | a,** Illustration of the Fermi-Dirac-like distribution of Δn and multilayer slabs. The gradient of blue shows the smooth transition from the birefringent region to isotropic region. **b,** Fitting parameter $\alpha$ determines how broad the interface is. **c,** The result of calculation for different values of $\alpha$. There is a significant decrease in the oscillation amplitude in birefringence if the interface is broader.

The result of the calculation shown in Fig. S3(c) for different values of $\alpha$. If the interface is broader, i.e. $\alpha$ is larger, there is a significant overall decrease in amplitude, indicating that a broadened distribution of Δn at the interface provides a better fit to the data. This broadened-interface model in combination with the experimental data demonstrates a novel technique for profiling depth dependence of the transition temperature.

## IV. Derivation of the nematic free energy

We start by discussing the symmetry properties of $Fe_{1/3}NbS_2$. The point group embedded in its space group ($P6_322$) is $D_6$, which is generated by a six-fold rotation $C_{6z}$ and a perpendicular twofold rotation $C_{2x}$. In space group $P6_322$ the sixfold rotation becomes a six-fold screw $S_{6z} = \{C_{6z}|00\frac{1}{2}\}$. As a result, the space group is generated by $S_{6z}$ and $\{C_{2x}|000\}$. Nematic order lowers the six-fold rotation $C_{6z}$ to twofold $C_{2z}$.

Since the magnetic ordering originates from the Fe sites, we focus on the Fe layers of the crystal structure. Two inequivalent Fe layers should be distinguished, which are exchanged under the screw $S_{6z}$. If we label the two layers $A$ and $B$ then we obtain an effective honeycomb geometry, where the $A, B$ layers play the role of the honeycomb sublattices. Introducing a layer index $\alpha = A, B$ a general spin configuration can be expressed in terms of Fourier components as

$$\mathbf{S}_\alpha(\mathbf{R}) = \sum_\mathbf{q} \mathbf{S}_\alpha(\mathbf{q}) \, e^{i\mathbf{q}\cdot\mathbf{R}}. \tag{10}$$

Here we consider magnetic ordering with three ordering vectors $\mathbf{Q}_{i=1,2,3}$ related by $2\pi/3$ rotations. In the case of stripe order, they correspond to the three $\Gamma - M$ directions of the hexagonal Brillouin zone:

$$\mathbf{Q}_1 = \frac{\pi}{\sqrt{3}}(\sqrt{3},1), \quad \mathbf{Q}_2 = \frac{\pi}{\sqrt{3}}(-\sqrt{3},1), \quad \mathbf{Q}_3 = \frac{\pi}{\sqrt{3}}(0,-2). \tag{11}$$

Magnetic ordering at these ordering vectors implies that only $\mathbf{S}_\alpha(\mathbf{Q}_i)$ and $\mathbf{S}_\alpha(-\mathbf{Q}_i)$ are nonzero. Since $\mathbf{Q}_i = -\mathbf{Q}_i$ modulo a reciprocal lattice vector, we only need to consider $\mathbf{S}_\alpha(\mathbf{Q}_i)$.

Clearly, $\mathbf{S}_A(\mathbf{Q}_i)$ and $\mathbf{S}_B(\mathbf{Q}_i)$ transform into each other under the sixfold screw $S_{6z}$. We can form irreducible magnetic order parameters by taking the sum and difference, which are respectively even and odd under the screw rotation. For concreteness, in what follows we focus on the even magnetic order parameters $\mathbf{L}_i$:

$$\mathbf{L}_i = \mathbf{S}_A(\mathbf{Q}_i) + \mathbf{S}_B(\mathbf{Q}_i), \tag{12}$$

but the results are the same also for the odd case. Symmetry constrains the magnetic free energy density to:

$$F = a \sum_i \mathbf{L}_i^2 + \frac{u}{2}\left(\sum_i \mathbf{L}_i^2\right)^2 - \frac{g}{2}[(\mathbf{L}_1^2 + \mathbf{L}_2^2 - 2\mathbf{L}_3^2)^2] \\ + \frac{w}{2}[(\mathbf{L}_1 \cdot \mathbf{L}_2)^2 + (\mathbf{L}_2 \cdot \mathbf{L}_3)^2 + (\mathbf{L}_1 \cdot \mathbf{L}_3)^2]. \tag{13}$$

Comparing to the expression (1) in the main text, here we have $u = 2v_0 + \frac{2v_1}{3}$, $g = \frac{v_1}{6}$, and $w = 2v_2$. A stripe ground state, in which only one of the $\mathbf{L}_i$ becomes non-zero, takes place when $g > 0$ and $w > -12g$. This state not only breaks spin-rotational symmetry, but also the $C_{6z}$ rotational symmetry, since one among the three $\mathbf{Q}_i$ wave-vectors is chosen. Within a mean-field approach, as long as $u$ is large enough, both the magnetic (spin-rotational symmetry-breaking) and nematic ($C_{6z}$ to $C_{2z}$ symmetry-breaking) transitions happen at the same temperature $T_0$ as a second-order transition when $a \propto T - T_0 < 0$.

Fluctuations, however, completely change this picture. Formally, fluctuations can be included self-consistently by decoupling the first two quartic terms in Eq. (13) via three Hubbard-Stratonovich fields:

$$\begin{aligned} \psi &= u(\mathbf{L}_1^2 + \mathbf{L}_2^2 + \mathbf{L}_3^2) \\ n_1 &= g(\mathbf{L}_1^2 + \mathbf{L}_2^2 - 2\mathbf{L}_3^2) \\ n_2 &= g\sqrt{3}(\mathbf{L}_1^2 - \mathbf{L}_2^2). \end{aligned} \tag{14}$$

Physically, $\psi$ is simply the amplitude of the magnetic fluctuations. As for $n_1$ and $n_2$, they transform as $d_{x^2-y^2}$ quadrupolar order and as $d_{xy}$ quadrupolar order. These are precisely the transformation properties of the two components of a generic two-dimensional electronic nematic order parameter. In terms of the $D_6$ point group, the vector $\mathbf{n} = (n_1, n_2)$ transforms as the two-dimensional $E_2$ irreducible representation. In terms of the Hubbard-Stratonovich fields in Eq. (14), the free energy density (13) is given by:

$$F[\mathbf{L}_i, \psi, \phi_i] = (a + \psi) \sum_i \mathbf{L}_i^2 - \frac{\psi^2}{2u} + \frac{n^2}{2g} - n_1(\mathbf{L}_1^2 + \mathbf{L}_2^2 - 2\mathbf{L}_3^2) - n_2\sqrt{3}(\mathbf{L}_1^2 - \mathbf{L}_2^2) \tag{15}$$

with $\phi^2 = \phi_1^2 + \phi_2^2$. The partition function is given by:

$$Z = \int \mathcal{D}\mathbf{L}_i \, \mathcal{D}\psi \, \mathcal{D}n_i \, \exp\left(-\int_q F[\mathbf{L}_i, \psi, \phi_i]/T_0\right). \tag{16}$$

Here, $\int_q \equiv \int \frac{d^d q}{(2\pi)^d}$ and fluctuations of the $\mathbf{L}_i$ fields are taken into account by including the gradient term in Eq. (15), yielding:

$$F[\mathbf{L}_i, \psi, \phi_i] = \sum_i \chi_{ii}^{-1}(\mathbf{q}) \, \mathbf{L}_{i,q} \, \mathbf{L}_{i,-q} - \frac{\psi^2}{2u} + \frac{n^2}{2g}. \tag{17}$$

The diagonal inverse magnetic susceptibility matrix $\chi_{ii}^{-1}$ is given by:

$$\chi_{ii}^{-1}(\mathbf{q}) = \begin{pmatrix} a + \psi + q^2 - n_1 - \sqrt{3}n_2 & 0 & 0 \\ 0 & a + \psi + q^2 - n_1 + \sqrt{3}n_2 & 0 \\ 0 & 0 & a + \psi + q^2 + 2n_1 \end{pmatrix} \tag{18}$$

In the paramagnetic state, magnetic fluctuations can be integrated out. It is convenient to change variables to $r \equiv \psi + a$; the resulting partition function becomes:

$$Z = \int \mathcal{D}r \, \mathcal{D}n_i \, \exp(-f_{\text{eff}}[r, n_i]), \tag{19}$$

where we absorbed the constant $T_0$ in the coupling constants $u$ and $g$ and derived the free energy:

$$f_{\text{eff}}[r, n_i] = \frac{N}{2} \sum_i \int_q \log(\chi_{ii}(\mathbf{q})) - \frac{(r - a)^2}{2u} + \frac{n^2}{2g}. \tag{20}$$

Here, $N$ is the number of components of the $\mathbf{L}_i$ order parameters. Instead of working with $n_1$ and $n_2$, it is convenient to parametrize $n_1 = n \cos 2\theta$ and $n_2 = n \sin 2\theta$. The magnetic susceptibility matrix becomes, in this parametrization:

$$\chi_{ii}^{-1}(\mathbf{q}) = \begin{pmatrix} r + q^2 - 2n \cos\left(2\theta - \frac{\pi}{3}\right) & 0 & 0 \\ 0 & r + q^2 - 2n \cos\left(2\theta - \frac{5\pi}{3}\right) & 0 \\ 0 & 0 & r + q^2 - 2n \cos(2\theta - \pi) \end{pmatrix}. \quad (21)$$

Up to this point, all results are exact. To proceed, we must perform some approximation. It is instructive to expand the free energy in powers of $n$ only. We find:

$$\frac{f_{\text{eff}}}{N} = \frac{n^2}{2}\left[\frac{1}{g} - \int_q \frac{3}{(r+q^2)^2}\right] + \left[\int_q \frac{1}{(r+q^2)^3}\right] n^3 \cos 6\theta - \left[\int_q \frac{9}{(r+q^2)^4}\right] n^4 + (\cdots), \quad (22)$$

where, as customary, we renormalized $(u, g) \to (u, g)/N$. This is precisely the free energy for the 3-state Potts model discussed in the main text. The key point is that, from Eq. (21), $r$ can be interpreted as the inverse squared magnetic correlation length $\xi^{-2}$ in the absence of Potts-nematic order. Thus, $r$ must vanish at the bare magnetic transition. This implies that the quadratic coefficient of $n^2$ changes sign from positive (when $r \to \infty$) to negative (when $r \to 0$) before the magnetic transition is reached. As a result, Potts-nematic order appears already in the paramagnetic phase. Note that, because of the positive cubic coefficient, the free energy is minimized by one of the three values $\theta = \pi/6, \pi/2, 5\pi/6$. Note also that in principle $r$ can also be a function of $n$, which will alter the quartic coefficient in front of $n^4$.

The existence of the cubic coefficient implies that, in three dimensions, the Potts-nematic transition is mean-field and first-order. This opens up the possibility of a simultaneous first-order Potts-nematic and magnetic transition, depending on how large the jump in $n$ is. This can be seen directly from the inverse susceptibility, Eq. (21): it is clear that the largest component of the $q = 0$ magnetic susceptibility is given by $\chi_{ii} = 1/(r - 2n)$, where we used the fact that $\theta$ assumes one of the values $\theta = $ $\pi/6, \pi/2, 5\pi/6$. Now, $n$ jumps by $\Delta n$ for a finite value of $r = r_c$. If however $\Delta n > r_c/2$, this implies a diverging renormalized magnetic susceptibility, and thus a simultaneous first-order magnetic transition.

Similar results hold also if the three ordering vectors $\mathbf{Q}_i$ are incommensurate and aligned along the $\Gamma - K$ and $\Gamma - K'$ directions. For instance, a zigzag phase is obtained when the ordering vectors are given by

$$\mathbf{Q}_1 = \frac{\pi}{2}(1, \sqrt{3}), \quad \mathbf{Q}_2 = \frac{\pi}{2}(-2, 0), \quad \mathbf{Q}_3 = \frac{\pi}{2}(1, -\sqrt{3}). \tag{23}$$

These ordering vectors are located on the $\Gamma - K$ and $\Gamma - K'$ high symmetry lines, are related by threefold rotation, and have the property that $\mathbf{Q}_i \neq -\mathbf{Q}_i$. The latter implies that, in contrast to Eq. (12), the corresponding order parameters $\mathbf{L}_i$ are complex. As a result, the nematic order parameter is now given by:

$$\begin{aligned} n_1 &\propto (|\mathbf{L}_1|^2 + |\mathbf{L}_2|^2 - 2|\mathbf{L}_3|^2) \\ n_2 &\propto \sqrt{3} \, (|\mathbf{L}_1|^2 - |\mathbf{L}_2|^2) \end{aligned} \tag{24}$$

## V. Domain Repopulation

We model the repopulation of the $\mathbb{Z}_3$-nematic domains under uniaxial strain, demonstrated in the main text Fig. 5, with a free energy, $F$, that contains two terms,

$$F = [(x_1 - x_2)^2 + (x_2 - x_3)^2 + (x_3 - x_1)^2] - b \sum_{n=1}^{3} x_n \epsilon \cos 2(\theta - \theta_n). \tag{25}$$

The first term favors equal domain populations whereas the second term breaks the degeneracy of the three domains. Here $x_n$ is the areal fraction of the $n^{\text{th}}$ domain, $b$ is the strain-nematic coupling parameter, $\epsilon$ is the amplitude of the uniaxial strain, $\theta - \theta_n$ is angle of the strain with respect to the director of domain $n$, and $\theta_n = 2\pi n/3$. Using the constraint $\Sigma x_n = 1$ and linearizing with respect to $\delta_n \equiv x_i - 1/3$, we find that $F$ is minimized for $\delta_n = (3b\epsilon/4) \cos 2(\theta - \theta_n)$. The model predicts that for tensile strain along a bond the domain aligned with the strain grows according to $\delta_0 = 3b\epsilon/4$, while the other two shrink, $\delta_1 = \delta_2 = -3b\epsilon/8$, and the signs of the $\delta_n$ are reversed for tensile strain perpendicular to the bond direction.

The domain repopulation in the $\mathbb{Z}_3$ nematic enables the global symmetry axes to continuously follow the direction of in-plane $\boldsymbol{B}$, $\boldsymbol{J}$, or uniaxial strain fields, despite the locking of the Neel vector perpendicular to the plane. As an indicator of the global symmetry, we consider the resistivity anisotropy tensor, $\overleftrightarrow{\boldsymbol{\rho}}_n \equiv R_{\theta_n} \Delta\rho \sigma_z$ where $R_{\theta_n}$ is the rotation operator and $\Delta\rho$ and $\sigma_z$ are the resistivity anisotropy and the Pauli $z$ matrix, respectively. We assume that the global symmetry axes align with the spatial average tensor, $\langle\overleftrightarrow{\boldsymbol{\rho}}\rangle = \Delta\rho \sum \delta_n, \overleftrightarrow{\boldsymbol{\rho}}_n$. At equilibrium, all three domains are equally populated ($x_i = 1/3$) and $\langle\overleftrightarrow{\boldsymbol{\rho}}\rangle$ is a null tensor, corresponding to a globally isotropic system. However, an external field applied at an angle $\theta$ unbalances the domain population, in which case,

$$F\langle\overleftrightarrow{\boldsymbol{\rho}}\rangle \propto \Delta\rho \begin{bmatrix} \cos 2\theta & \sin 2\theta \\ -\sin 2\theta & \cos 2\theta \end{bmatrix}, \tag{26}$$

which shows that rotational symmetry is broken and the global symmetry axis continuously follows the direction of the perturbation. This is in stark contrast with the $\mathbb{Z}_2$-Ising-nematic system where the global anisotropy axes are the locked to the crystal axes, regardless of the direction of an external perturbation.

## References

1. Doyle, S. *et al.* Tunable Giant Exchange Bias in an Intercalated Transition Metal Dichalcogenide. Preprint at https://arxiv.org/abs/1904.05872 (2019).
2. Fan, S. *et al.* Electronic chirality in the metallic ferromagnet $Fe_{1/3}TaS_2$. *Phys. Rev. B* **96**, 1–6 (2017).
3. Guenther, R.D. Modern Optics. Wiley, (1990).